# GS-QA: Comprehensive Quality Assessment Benchmark for Gaussian Splatting View Synthesis

Pedro Martin, António Rodrigues, João Ascenso, and Maria Paula Queluz
Instituto de Telecomunicações, Instituto Superior Técnico, University of Lisbon, 1049-001 Lisbon, Portugal
{pedro.martin, antonio.rodrigues, joao.ascenso, paula.queluz}@lx.it.pt

***Abstract*—Gaussian Splatting (GS) offers a promising alternative to Neural Radiance Fields (NeRF) for real-time 3D scene rendering. Using a set of 3D Gaussians to represent complex geometry and appearance, GS achieves faster rendering times and reduced memory consumption compared to the neural network approach used in NeRF. However, quality assessment of GS-generated static content is not yet explored in-depth. This paper describes a subjective quality assessment study that aims to evaluate synthesized videos obtained with several static GS state-of-the-art methods. The methods were applied to diverse visual scenes, covering both 360º and forward-facing (FF) camera trajectories. Moreover, the performance of 18 objective quality metrics was analyzed using the scores resulting from the subjective study, providing insights into their strengths, limitations, and alignment with human perception. All videos and scores are made available providing a comprehensive database that can be used as benchmark on GS view synthesis and objective quality metrics.**

## I. Introduction

In recent years, 3D visual representations have become increasingly popular across several domains, including virtual reality, gaming, medical imaging, and architecture [1]. These representations offer immersive and interactive experiences, enabling users to perceive depth and spatial relationships more naturally compared to traditional 2D images. The latest advancement in 3D scene representation, Gaussian Splatting (GS) [2], offers a more efficient alternative to Neural Radiance Fields (NeRF) [3], by using anisotropic 3D Gaussians, enabling real-time rendering with exceptional computational efficiency. In contrast to NeRF's reliance on the computationally intensive ray marching, GS employs a tile-based rasterizer to achieve real-time rendering. This approach delivers competitive visual quality and significantly lower training and synthesis times compared to most NeRF methods. Moreover, GS methods do not rely on neural networks and avoid computations in the empty 3D space.

Despite its many advantages, views generated by a GS model often exhibit artifacts, such as floaters and geometric distortions, which can significantly affect visual quality. The lack of a comprehensive subjective evaluation study for GS-generated static content highlights a critical gap in the literature. Moreover, while objective quality metrics such as PSNR, SSIM [4], and LPIPS [5] are commonly used to assess the quality of GS-generated content [1], their correlation with human perception remains insufficiently studied. Considering these limitations, the main objective of this work is to conduct a comprehensive subjective assessment campaign to perceptually compare the performance of several static GS methods and identify the most accurate objective quality metrics for evaluating GS synthesized views. The key contributions of this work are:

- **Subjective assessment campaign:** A subjective study was conducted with GS synthesized views, providing perceptual quality assessment of seven state-of-the-art GS-based methods, from the original work, 3D Gaussian Splatting (3DGS) [2], to the more recent Scaffold-GS [6]. Eight real-world static scenes were used and both 360º and forward-facing (FF) camera trajectories were considered to obtain the GS models. By applying view synthesis to the 3D models using the defined trajectories, videos are obtained and subsequently evaluated through a Double Stimulus Continuous Quality Scale (DSCQS) subjective test conducted in a semi-controlled laboratory environment.
- **Performance evaluation of objective metrics:** A thorough analysis of several objective quality metrics was conducted using the videos shown to the participants in the subjective assessment test, and the resulting perceptual scores. This evaluation offers insights into the metrics performance, highlights limitations in addressing GS-specific artifacts and distortions, and identifies the metrics most effective for assessing GS view synthesis quality, thereby guiding future developments.

The original and synthesized videos, along with the corresponding quality scores, are publicly available in the GS Quality Assessment (GS-QA) database[1].

The rest of this paper is organized as follows: Section II reviews related work. Section III introduces the GS-QA framework and the selected GS methods. Section IV describes the subjective assessment test, including datasets, methodology, and results analysis. Section V evaluates objective quality metrics, focusing on their correlation with human perception. Section VI concludes the paper with a discussion of key insights and their potential impact.

## II. Related Work

Several prior works have explored quality assessment of view synthesis techniques generated by NeRF methods. NeRF-QA [7] conducted the first subjective quality assessment on NeRF-synthesized videos, using a DSCQS methodology with videos generated from several NeRF methods. It established a foundational benchmark for understanding human perception of NeRF rendering quality. NVS-QA [8] expanded on NeRF-QA by extending the subjective study with a larger and more diverse selection of visual scenes. Additionally, it introduced a comprehensive evaluation of objective quality metrics, assessing their effectiveness in capturing NeRF-specific artifacts and distortions. E-NeRF-QA [9] focused on explicit NeRF models, which employ structured representations (e.g., voxel grids) for

This work is funded by FCT/MECI through national funds and when applicable co-funded EU funds under UID/50008 and PTDC/EEI-COM/7775/2020.

[1]github.com/pedrogcmartin/GS-QA

improved rendering efficiency and storage optimization. This work primarily examined the impact of lossy compression on perceptual quality, using a Double Stimulus Impairment Scale (DSIS) subjective study to analyze compression-induced degradations. Similar to NVS-QA, [10] further explored NeRF quality assessment by conducting a perceptual quality evaluation (adopting a pairwise comparison methodology) on NeRF-synthesized images, alongside an objective quality metric performance analysis.

The quality assessment of GS-generated content has been relatively underexplored, with only two studies — [11] and [12] — addressing this topic. In [11], a graph-based GS data compression method was proposed and evaluated using DSIS to assess compression artifacts in the GS-synthesized views. In [12] the authors conducted a quality assessment study comparing GS-based and NeRF-based methods, evaluating two methods from each category. They focused solely on dynamic scenes (featuring moving objects) and used the Subjective Assessment Methodology for Video Quality (SAMVIQ) to evaluate rendering quality. Moreover, [12] only considers two basic GS methods and lacks statistical significance test analysis to validate its findings.

Despite the availability of numerous GS methods, the quality assessment of GS static scenes has not yet been explored in previous studies. Moreover, many of the recent methods target specifically static scenes, which remain highly popular and widely applicable in immersive content applications. This study, therefore, directly addresses the quality assessment of GS-generated views, with particular attention to the rendering artifacts introduced by these methods. It also includes Mip-NeRF 360 [13], a NeRF-based method recognized as a top performer in visual quality [8]. The subjective assessment methodology allows to understand the performance of many relevant and recent GS methods. Finally, the study evaluates the performance of several objective metrics for this GS generated content.

## III. GS View Synthesis Framework

This section provides a short description of the selected GS methods and outlines the framework employed for the production of GS-based videos using novel views/trajectories.

### A. *Selected GS Methods for Evaluation*

From the increasing number of GS methods proposed in the literature, six were selected for performance evaluation alongside the seminal 3DGS [2] work. The selection criteria emphasized performance, popularity, and the implementation availability. The selected GS methods are described in chronological order based on their publication dates:

- **3D Gaussian Splatting (3DGS) [2]:** Proposes a novel explicit radiance field representation using anisotropic 3D Gaussians to model a scene's radiance field. Each Gaussian is parameterized by its position, rotation, scale, opacity, and Spherical Harmonics (SH) coefficients, enabling efficient view-dependent color rendering. The method employs a tile-based rasterizer and an adaptive density control mechanism that dynamically adjusts the number and distribution of Gaussians. This groundbreaking approach enables high-quality novel view synthesis while ensuring considerably faster training and rendering regarding NeRF methods.
- **Mip-Splatting [14]:** Proposes a solution for aliasing and high-frequency artifacts that occur in 3DGS rendering when changing sampling rates (e.g., zooming in/out). It introduces two key improvements: a 3D smoothing filter to suppress high-frequency artifacts when zooming in, and a 2D MIP filter that replaces traditional dilation with a Mipmap-like approach, for alias-free rendering at different scales. These modifications improve visual quality and robustness across varying resolutions while maintaining real-time performance.
- **LightGaussian [15]:** Proposes a Gaussian pruning and recovery strategy to reduce redundancy while preserving visual quality. A distillation process is used to obtain a more compact representation, reducing SH coefficients to a lower degree. Furthermore, vector quantization is applied on the Gaussian parameters reducing precision for less significant features. These techniques reduce the model size and boosts the speed of 3DGS.
- **Scaffold-GS [6]:** Proposes a structured approach by using anchor points to distribute local Gaussians. These anchors are placed in a sparse voxel grid and adapt dynamically based on viewing direction and distance, allowing efficient rendering with fewer primitives. Anchor growing and pruning strategies reduce redundant Gaussians and enhance rendering quality, particularly in scenes with diverse levels of detail (LOD) and view-dependent observations.
- **Efficient Accelerated 3D Gaussians with Lightweight Encodings (EAGLES) [16]:** Proposes quantized embeddings to compress the color and rotation parameters of each Gaussian, decreasing the overall memory storage. Additionally, it employs a coarse-to-fine training strategy to enhance optimization stability and convergence speed. A pruning stage is also incorporated to remove redundant Gaussians. Collectively, these strategies enable EAGLES to achieve faster training speeds than 3DGS, while maintaining high-quality results.
- **Self-Organizing Gaussians (SOG) [17]**: Proposes a method to organize Gaussian parameters into a 2D grid enabling efficient use of image codecs. A parallel linear assignment sorting algorithm arranges Gaussians efficiently, preserving their neighborhood structure, thus improving compression performance. By enforcing local smoothness between the sorted parameters in the 2D grid during training, it further enhances rendering quality.
- **Octree-GS [18]**: Proposes a LOD structured approach that organizes Gaussians within an octree structure, enabling adaptive rendering for large scenes with complex details. The method dynamically selects the appropriate LOD based on the camera view, achieving consistent real-time rendering performance across scenes of varying complexity.

All methods were executed on two NVIDIA GeForce RTX 4090 GPUs, except for SOG, which was specifically developed for a single GPU.

### B. *Framework Overview*

The framework takes video sequence(s) as input, which are split into training images, used to train GS methods, and a reference video, which serves as a baseline for conducting subjective and objective quality assessment studies. The camera trajectory of the reference video is also calculated and used for view synthesis, i.e., to generate the degraded (or distorted) video, ensuring the same camera trajectory in both video sequences (reference and synthesized). Fig. 1 illustrates the GS-QA framework pipeline, which is detailed below for the 3DGS case:

- **Train/test split:** The input video(s) are split into training images and a reference video. Different trajectories are used for training and test, which overlap in terms of scene coverage. In this work, the trajectories and the sampling of frames defined in [10] were used. Camera lens distortion is also corrected as pre-processing for the next step.
- **Pose estimation:** A reliable structure-from-motion (SfM) algorithm, COLMAP [19], [20], is employed to estimate camera poses. Pose estimation is performed for the entire scene and outputs both training and reference poses.
- **GS training:** The radiance field of a scene is parameterized using anisotropic 3D Gaussians. The Point Cloud (PC) generated by the pose estimation process is used to initialize a set of 3D Gaussians for the next step. However, the number of 3D Gaussians and the associated parameters are optimized in this step. Using a tile-based rasterizer, the Gaussians are projected onto 2D image planes considering the training poses. Stochastic gradient descent is then applied iteratively to find the set of parameters values that minimize the difference between the synthesized and training images. Furthermore, various strategies, including adaptive density control, anchor-based distributions, or hierarchical structures, may be employed to dynamically adjust the number and spatial distribution of Gaussians, enhancing detail where needed while reducing redundancies.
- **View synthesis:** The view synthesis process starts by projecting the 3D Gaussians onto the corresponding 2D image plane. This plane is defined according to the reference poses (test trajectory). Each Gaussian's position has a projected location on the image plane, while its rotation and scale define its anisotropic shape and orientation in the 2D plane. To enable parallel processing, the image plane is divided into smaller tiles allowing to render multiple parts of the image simultaneously. Each 3D Gaussian is evaluated to determine which tiles it intersects and thus only the relevant areas of the image are processed for each Gaussian, making the rendering more efficient. The Gaussians that intersect with a tile are "splatted" onto that tile, meaning their parameters contribute to the pixels in that tile. The contributions of all splats are combined to form the final 2D view of the scene based on their opacities. This blending process ensures a smooth, photorealistic reconstruction of the scene.

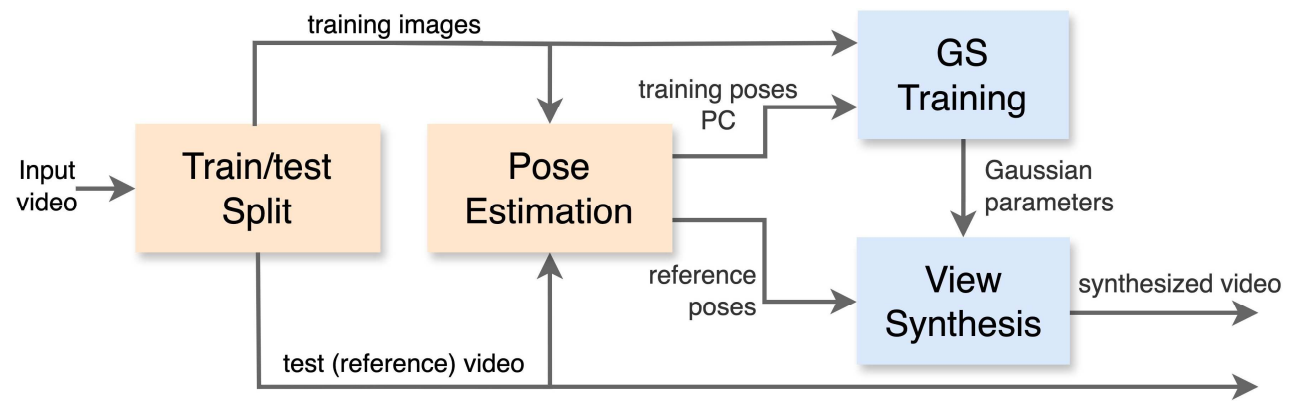

Fig. 1. GS-QA pipeline for synthesized and reference video creation.

## IV. Subjective Assessment Experiment

This section describes the subjective assessment study of GS-based synthesized videos and provides an analysis of the results. Subjective assessment is the ultimate way of evaluating quality, as it directly captures human perception.

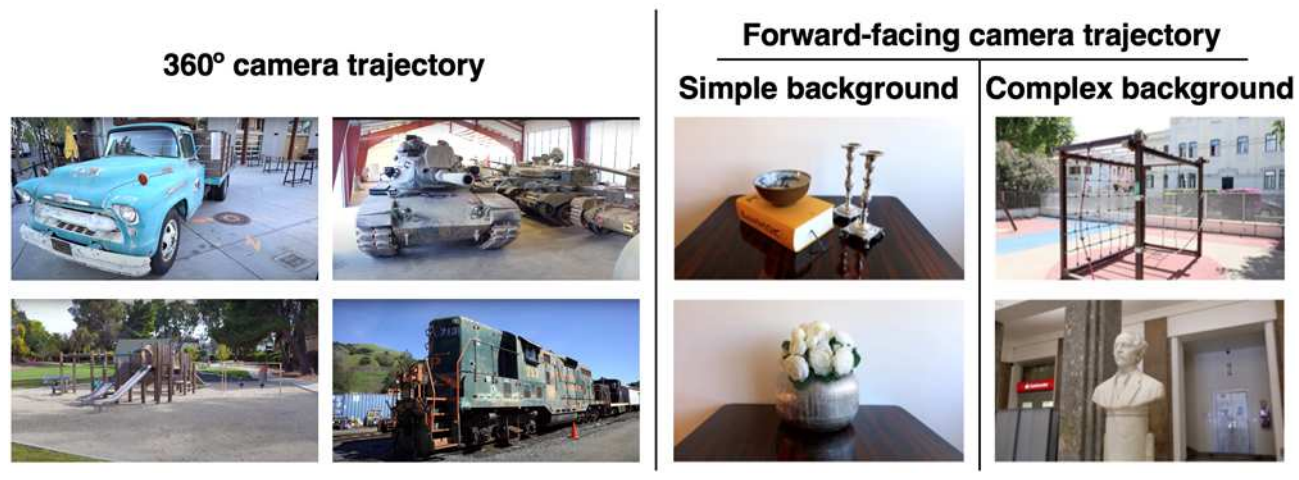

Fig. 2. Selected scenes from the T&T and IST/IT datasets.

### *A. Experimental Conditions*

For evaluation, two datasets were chosen: Tanks and Temples (T&T) [21], [22] and IST/IT [8] (cf. Fig. 2). Both datasets contain real-world captures using two camera trajectories: 360º, where the camera orbits a region of interest, facing inward, and FF, where the camera moves within a 2D grid, facing straight toward the region of interest. Four visual scenes were selected from each dataset: M60, Playground, Train, and Truck, from T&T dataset, and Antique, Flowers, Playground2, and Statue from IST/IT dataset. All the eight scenes feature complex backgrounds, except for Antique and Flowers, which depict a simple background (namely, a white wall). The T&T scenes have 277, 275, 258, and 226 training images, respectively, with spatial resolutions of 1007×546, 1008×548, 982×546, and 980×546 pixels. The IST/IT scenes have 251, 377, 291 and 228 training images, respectively, all with a spatial resolution of 1920×1080 pixels. The reference videos have a duration of 10 s and a temporal resolution of 24 fps (for M60), 25 fps (for all the IST/IT scenes), and 30 fps (for the remaining T&T scenes). By applying the GS-QA framework on the selected GS methods (presented in Section III.A) with the selected visual scenes, a total of 64 synthesized videos were generated, along with eight reference videos.

The DSCQS methodology, recommended in ITU-R BT.500-15 [23], was chosen to subjectively assess the quality of radiance field-based content. This method is particularly useful for cases where the processed video may surpass the reference in quality, a rare but possible scenario with radiance field-based content [8]. A FullHD monitor, with a spatial resolution of 1920×1080 pixels, was used for subjective assessment under semi-controlled conditions, with a viewing distance of three times the monitor's height. The final synthesized videos and their respective references were cropped to a spatial resolution of 928×522 pixels and displayed side-by-side. The IST/IT scenes were downsampled by a factor of two before cropping. A total of 22 non-expert-viewers, comprising 14 males and 8 females, aged between 18 and 33 years, participated in the subjective assessment campaign.

Before the start of each test session, participants were informed about the test objective. A training session followed, allowing them to familiarize with the evaluation procedure using video sequences labeled with suggested scores. Once the training was completed, participants proceeded to the test session. In each trial, they were shown two unlabeled videos: a reference video and a synthesized video generated by one of the selected GS methods. The synthesized video always follows the same trajectory as the reference video. The reference video's position (left or right) and the video pairs were randomized in each trial. Participants evaluated both stimuli using a continuous slider labeled with five quality levels (Bad, Poor, Fair, Good, and Excellent). The assigned scores were registered on a scale from 1 (Bad) to 5 (Excellent), with two decimal places.

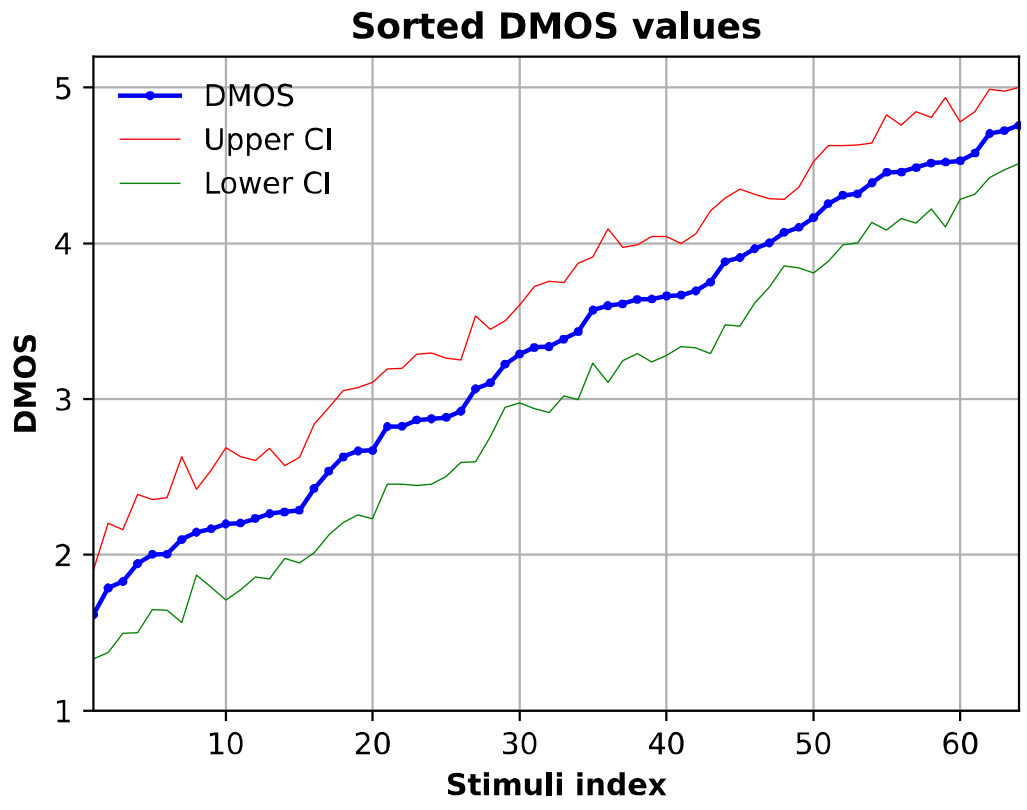

Fig. 3. DMOS values of the subjective study stimuli with 95% confidence intervals.

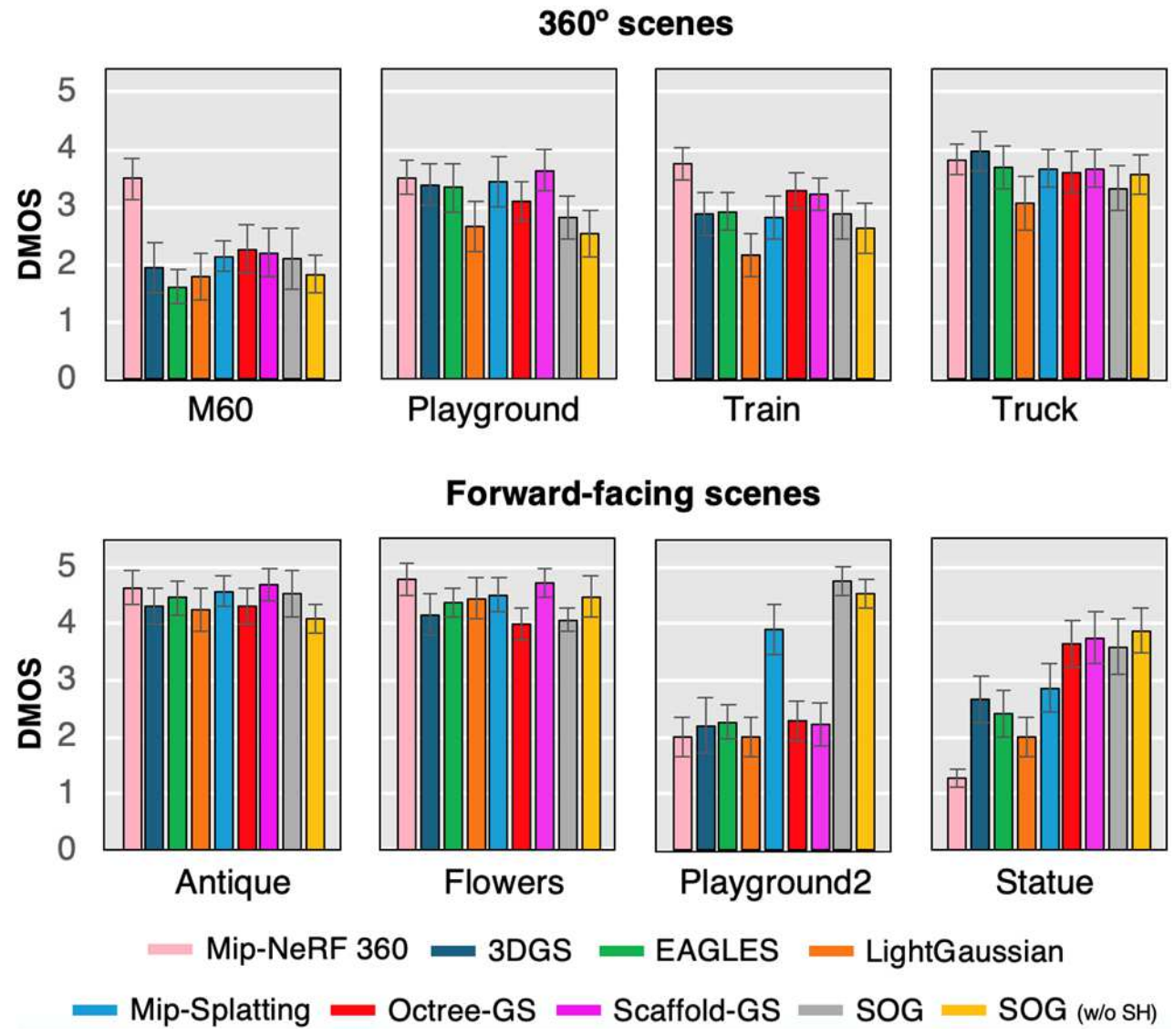

Fig. 4. DMOS values of the synthesized scenes per GS method with 95% confidence intervals.

A total of 64 pairs of stimuli were evaluated, consisting of four 360º scenes and four FF scenes generated by the considered methods. Following the test session, outliers were identified and removed using both Kurtosis-based and correlation-based post-screening methods described in ITU-R BT.500-15 [23], resulting in the exclusion of scores from two participants. Next, Mean Opinion Scores (MOS) were calculated for both the synthesized and reference video sequences by averaging the assigned scores from all valid participants. The resulting scores were processed in accordance with ITU-T P.910 [24] to derive Differential MOS (DMOS) values for each pair of evaluated videos (where higher DMOS values indicate better quality).

### B. Experimental Results

The DMOS scores, sorted in ascending order, along with their corresponding 95% confidence intervals, are presented in Fig. 3. As shown, the entire quality range is represented, which was expected considering the selected methodology and GS methods. Fig. 4 illustrates the DMOS scores by scene and GS method. The results from Mip-NeRF 360 [13], available in NVS-QA [8], are also included to evaluate how well the GS methods compare to the best NeRF-based method in terms of perceptual quality. Note that the same subjective test conditions and videos were used for both GS and NeRF quality assessment studies. Table I shows the average quality, speed, and size of all methods under evaluation. The average size for each method represents the mean (i.e., across all scenes) file size of the respective model parameters after the training process is completed. As shown, the fastest method for training and synthesis is LightGaussian, and the lowest model size is obtained for the SOG method. From both Fig. 4 and Table I, the following conclusions are taken:

- **NeRF vs. 3DGS approaches:** The synthesis quality of GS methods typically does not exceed the quality obtained with Mip-NeRF 360, except in FF scenes with complex backgrounds (Playground2 and Statue), where Mip-NeRF 360 has poor performance [8]. These findings are consistent with the claims in the seminal 3DGS work [2], which states that the primary goal of 3DGS is to enhance real-time rendering efficiency, not necessarily to improve visual quality. As shown in Table I, GS methods exhibited significantly faster training and rendering speeds than Mip-NeRF 360, by approximately two orders of magnitude, highlighting their computational efficiency advantage. This difference primarily stems from the tile-based rasterizer used in GS, instead of the ray marching technique employed by NeRF, for view synthesis.
- **GS methods evaluation:** Among the GS methods, Scaffold-GS achieved the highest rendering quality for 360º scenes (cf. Table I), significantly outperforming the other GS approaches, though at the expense of being the slowest GS method. For FF scenes, SOG emerged as a top performer in terms of both rendering quality and model size. Interestingly, the removal of spherical harmonics in SOG proved to be beneficial in certain scenes (Flowers, Statue, and Truck), which is compensated by using a larger number of Gaussians to capture fine scene details. These findings align with prior observations in the SOG work [17]. Both Scaffold-GS and SOG demonstrated significantly higher quality performance compared to other methods, across all scenes. Moreover, LightGaussian offered the best trade-off between speed and storage efficiency, though at the cost of reduced visual quality.

TABLE I. QUALITY, SPEED, AND STORAGE PERFORMANCES OF THE SELECTED GS METHODS

| Method | Average DMOS | | | Average Speed [min] | | Average Size [MB] |
|---|---|---|---|---|---|---|
| | 360 | FF | All | Train. | Syn. | |
| Mip-NeRF 360 | **3.64** | 3.18 | 3.41 | ~1800 | 23.70 | 108.1 |
| 3DGS | 3.04 | 3.33 | 3.19 | 10.00 | 1.88 | 258.3 |
| Mip-Splatting | 3.02 | 3.97 | 3.49 | 16.63 | 2.38 | 343.1 |
| LightGaussian | 2.42 | 3.18 | 2.80 | **0.40** | **0.38** | 18.9 |
| Scaffold-GS | 3.18 | 3.85 | **3.52** | 24.88 | 3.38 | 156.6 |
| EAGLES | 2.89 | 3.39 | 3.14 | 9.43 | 0.77 | 199.6 |
| SOG* | 2.78 | 4.24 | 3.51 | 18.25 | 2.50 | 13.3 |
| SOG* (w/o SH) | 2.64 | **4.25** | 3.45 | 17.50 | 2.75 | **5.6** |
| Octree-GS | 3.07 | 3.56 | 3.31 | 19.52 | 3.40 | 48.1 |

*Run in a single GPU.

TABLE II. STATISTICAL TEST RESULTS FOR ALL GS METHODS AND TYPES OF SCENES.

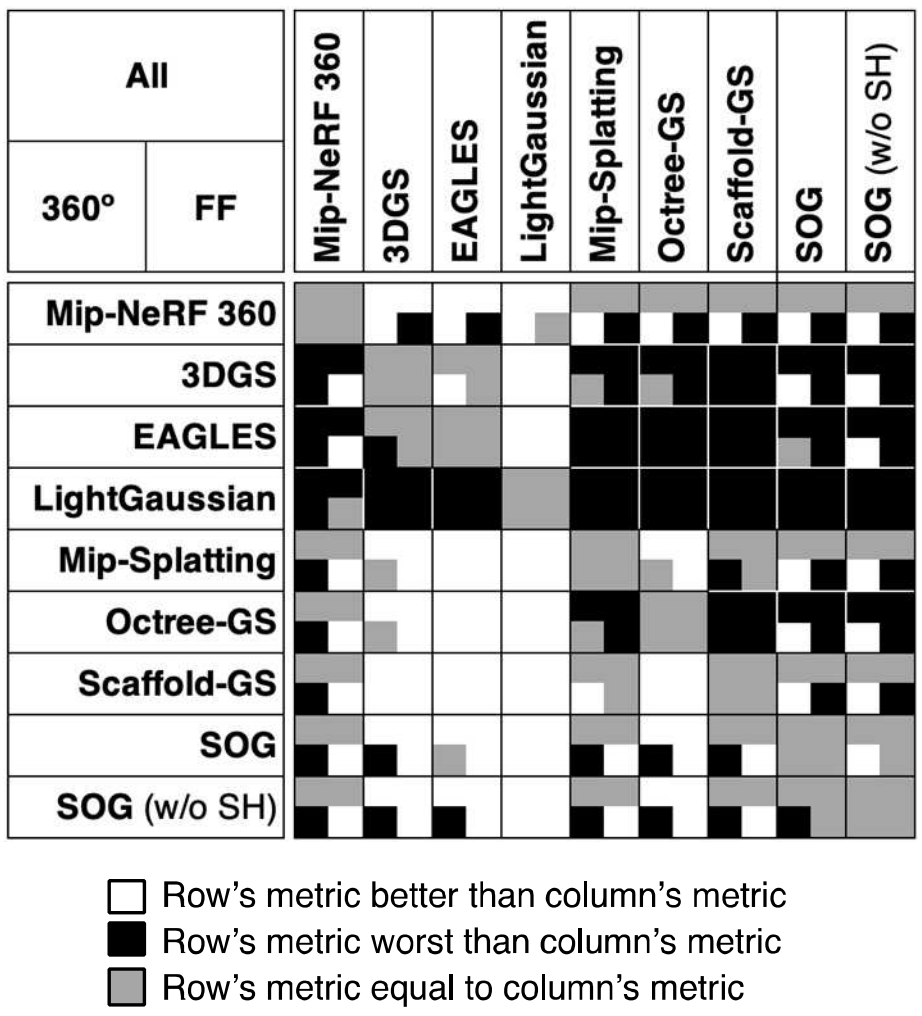

- **Scene type:** GS methods achieved, on average, 29% higher quality scores in FF scenes compared to 360º scenes; among the FF scenes, Antique and Flowers obtained the highest DMOS values (cf. Fig. 4). However, the DMOS scores in FF scenes vary more across different scenes compared to 360º scenes. The lower DMOS of scenes like Playground2 and Statue stem from the increased complexity of backgrounds, which pose greater challenges for accurate representation. In contrast, simpler backgrounds, as seen in Antique and Flowers, reduce the likelihood of hallucinated artifacts such as floaters, contributing to higher visual quality.

The performance differences between the methods were further analyzed for statistical significance using the procedure outlined in [25]. A two-sided paired sample t-test was conducted for all possible pair of methods. Prior to the t-test, assumptions of normality and homogeneity of variance were verified. The Shapiro-Wilk test was used to assess normality for sample sizes below or equal to 50, while the Kolmogorov-Smirnov test was applied for larger samples. Variance homogeneity was evaluated using Levene's test. Optimal significance thresholds were defined using a decision-theoretic approach [26]. Table II summarizes the results of the GS methods' significance tests. In this Ta' ' white cells indicate that the method in the row significa outperformed the method in the column, black cells indi the opposite, and gray cells denote no significant differe Each cell contains results for 360º scenes (bottom-left), scenes (bottom-right), and all scenes (top). The statist analysis of the methods' performances confirms Mip-N 360 as the top performer for 360º scenes, followed Scaffold-GS, while SOG demonstrates the highest qualit FF scenes.

## V. QUALITY METRICS PERFORMANCE ASSESSMENT

This section evaluates the performance of several objec quality metrics, providing an in-depth analysis of their res and offering insights into their suitability for GS-gener content.

### A. Selected Metrics for Evaluation

A comprehensive range (18 in total) of objective f reference quality assessment metrics were selected evaluation. They can be grouped in two main categories: *i)* metrics following a classical approach, including FSIM [27], FVVDP [38], GMSD [28], IW-SSIM [29], MAD [30], MSE-RGB, MS-SSIM [31], NLPD [32], PSNR-HVS [33], PSNR-Y, PSNR-YUV, SSIM [4], VIF [34], and VSI [35]; *ii)* learning-based metrics that are end-to-end trained, such as DISTS [36], LPIPS [5], ST-LPIPS [37], and VMAF [39]. Only FVVDP and VMAF were developed for video. For the quality image metrics, the frame-based quality scores were averaged to compute the overall video quality score.

### B. Experimental Results

To evaluate the performance of the objective metrics, the Pearson Linear Correlation Coefficient (PLCC), Spearman Rank Order Correlation Coefficient (SROCC), and Perceptually Weighted Rank Correlation (PWRC) were employed. These correlation metrics compare the predicted (by the objective metrics) DMOS values against the ground-truth values derived from subjective tests. Since subjective quality scores often saturate at the extremes of rating scales, a non-linear mapping between objective quality metrics and DMOS values was applied using cubic regression, in accordance with the ITU-T P.1401 recommendation [40]. In contrast to conventional ranking metrics, such as SROCC, PWRC focuses on the ranking accuracy of high-quality images and accounts for the inherent uncertainty in subjective evaluations of images with similar quality [41]. PWRC introduces the concept of a sensory threshold, defining the point at which viewers can discern differences between images. Table III presents the PLCC, SROCC, and PWRC results (at the sensory threshold of zero, meaning maximum sensitivity). In Table III, a heatmap is applied to the coefficient values, with the highest correlated metrics highlighted in blue and the lowest in red. The following conclusions can be drawn from these results:

- **Metrics performance:** According to Table III, similarity-based metrics consistently show strong performance. FSIM and MS-SSIM stand out as the best metrics for assessing 360º scenes, while ST-LPIPS outperform the other metrics, across all three coefficient indicators, in FF scenes. Notably, ST-LPIPS reached 0.9 PLCC and 0.8 in SROCC correlation in FF scenes. Similarity-based metrics evaluate key perceptual features: luminance, contrast, and structure which are crucial for 360º scenes where perceptual differences may be significant across views. MS-SSIM provides a multi-scale approach, and IW-SSIM uses content-weighted quality pooling. Conversely, FF scenes involve more restricted camera motion during acquisition, often resulting in the presence of subtle artifacts like pixel misalignments and geometric errors in the synthesized content. ST-LPIPS effectively handles these discrepancies, as it is tolerant to minor geometric variations while capturing high-level perceptual features, using deep neural networks.

TABLE III. PLCC, SROCC, AND PWRC PERFORMANCE RESULTS

| Metric | PLCC | | | SROCC | | | PWRC | | |
|---|---|---|---|---|---|---|---|---|---|
| | 360º | FF | All | 360º | FF | All | 360º | FF | All |
| FSIM | 0.95 | 0.76 | 0.81 | 0.94 | 0.67 | 0.81 | 0.81 | 0.41 | 0.67 |
| FVVDP | 0.79 | 0.75 | 0.80 | 0.81 | 0.68 | 0.80 | 0.61 | 0.58 | 0.59 |
| GMSD | 0.94 | 0.75 | 0.79 | 0.92 | 0.69 | 0.78 | 0.77 | 0.40 | 0.63 |
| IW-SSIM | 0.93 | 0.73 | 0.83 | 0.91 | 0.65 | 0.81 | 0.77 | 0.55 | 0.67 |
| MAD | 0.90 | 0.68 | 0.80 | 0.82 | 0.45 | 0.77 | 0.67 | 0.33 | 0.61 |
| MSE-RGB | 0.82 | 0.66 | 0.72 | 0.69 | 0.62 | 0.69 | 0.47 | 0.51 | 0.51 |
| MS-SSIM | 0.95 | 0.76 | 0.83 | 0.93 | 0.65 | 0.81 | 0.79 | 0.54 | 0.66 |
| NLPD | 0.94 | 0.74 | 0.80 | 0.92 | 0.66 | 0.74 | 0.78 | 0.43 | 0.60 |
| PSNR-HVS | 0.80 | 0.68 | 0.76 | 0.84 | 0.63 | 0.74 | 0.64 | 0.53 | 0.57 |
| PSNR-Y | 0.79 | 0.63 | 0.72 | 0.82 | 0.62 | 0.68 | 0.62 | 0.51 | 0.51 |
| PSNR-YUV | 0.79 | 0.61 | 0.68 | 0.68 | 0.53 | 0.64 | 0.46 | 0.42 | 0.46 |
| SSIM | 0.91 | 0.76 | 0.78 | 0.91 | 0.66 | 0.74 | 0.75 | 0.42 | 0.61 |
| VIF | 0.78 | 0.61 | 0.71 | 0.76 | 0.56 | 0.65 | 0.56 | 0.39 | 0.54 |
| VSI | 0.94 | 0.65 | 0.77 | 0.93 | 0.49 | 0.74 | 0.78 | 0.38 | 0.59 |
| DISTS | 0.94 | 0.68 | 0.80 | 0.93 | 0.50 | 0.78 | 0.79 | 0.36 | 0.62 |
| LPIPS | 0.75 | 0.81 | 0.73 | 0.76 | 0.69 | 0.72 | 0.57 | 0.58 | 0.56 |
| ST-LPIPS | 0.79 | 0.90 | 0.87 | 0.81 | 0.80 | 0.88 | 0.60 | 0.67 | 0.69 |
| VMAF | 0.83 | 0.69 | 0.80 | 0.76 | 0.64 | 0.78 | 0.61 | 0.52 | 0.61 |
| Averages | 0.86 | 0.71 | 0.78 | 0.84 | 0.62 | 0.75 | 0.67 | 0.47 | 0.59 |

- **Scene type:** On average, 360º scenes achieved higher correlation results than FF scenes, with both PLCC and SROCC indicators exceeding a correlation of 0.8 (cf. Table III). While FF scenes generally resulted in higher DMOS values compared to 360º scenes, they also exhibited greater DMOS variability across different scenes. Specifically, FF scenes with complex backgrounds have lower subjective scores (cf. Fig. 4), heavily influenced by visual artifacts such as floaters and inconsistent hallucinations in occluded areas. The selected objective quality metrics are not able to accurately assess such artifacts for several reasons: *i)* they are a relatively new type of artifacts in view synthesis and were not considered during the development of the metrics; *ii)* despite being highly noticeable to human viewers, they are not spatially dominant within the scene, often leading to a more critical subjective response than what the metrics capture; and *iii)* they lack temporal predominance, and since image quality metrics results are averaged across all synthesized images of a scene, the artifacts impact is smoothed out, reducing their influence on the overall quality score.
- **NVS-QA vs. GS-QA:** Compared to the NVS-QA dataset [8], the results obtained for GS-QA are different since the ranking order is not the same. The main reason is the distinct characteristics of the NeRF and GS view synthesis processes. NeRF-based methods are prone to spatial misalignments between synthesized and reference views. In such case, a disparity compensation step is often required to allow the quality metrics to have high performance. In contrast, GS-based methods exhibit fewer spatial misalignments, and thus disparity compensation does not need to be performed. Learning-based metrics such as DISTS and ST-LPIPS, known for their robustness to geometric distortions, also have higher performance in evaluating NeRF-generated content [8]. Similarity-based metrics such as FSIM and MS-SSIM outperformed others just for GS-based content assessment. Interestingly, ST-LPIPS stands out as a consistently effective metric across both types of methods.

The performance differences between the objective metrics were further analyzed for statistical significance using also the procedure outlined in [25] and described in Section IV.B. The two-sided paired sample t-test was conducted for all possible metric pairs, with prediction residuals for each metric as input. Prediction residuals were computed as the absolute differences between the predicted DMOS from each metric and the ground-truth subjective DMOS. The predicted DMOS was derived by applying a cubic regression to the objective metrics' scores [40]. All prediction residuals satisfied the t-test assumptions. Table IV summarizes the results of the metrics' significance tests. Notably, FSIM and MS-SSIM emerge as the only metrics that perform statistically significantly better than, or equal to, all other metrics across both 360º and overall scenes. Also, ST-LPIPS stands out as the only metric achieving statistically superior or equivalent performance across both FF and overall scenes.

TABLE IV. STATISTICAL TEST RESULTS FOR ALL OBJECTIVE METRICS AND TYPES OF SCENES.

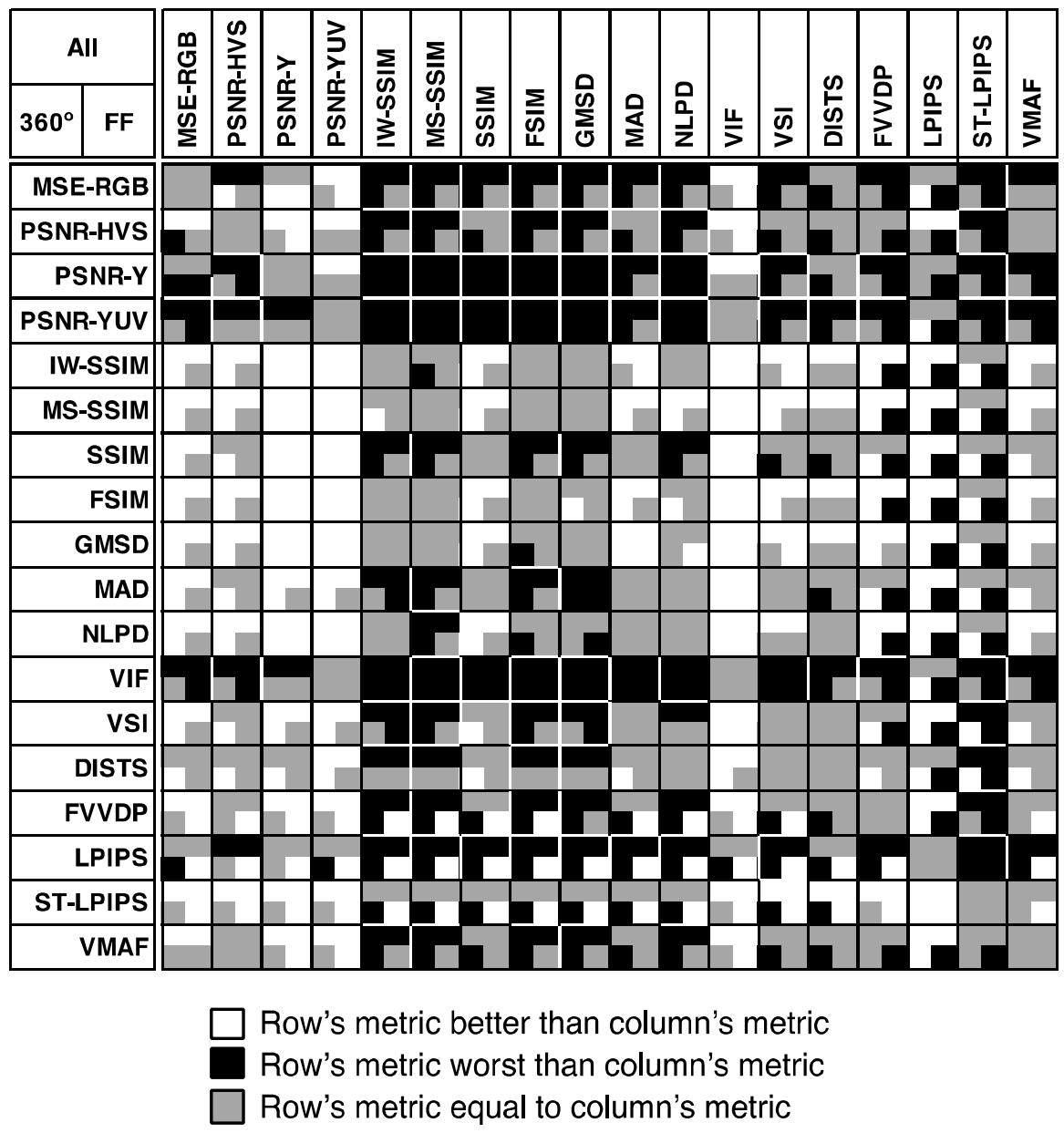

## VI. FINAL REMARKS

This paper introduces a comprehensive subjective quality assessment of GS methods. The visual quality of synthesized videos generated by several state-of-the-art GS methods was compared against the corresponding reference videos using the DSCQS test methodology, capturing human visual judgments. Additionally, objective quality metrics were evaluated for their effectiveness in assessing GS-generated content, offering valuable insights into their performance. A statistical analysis was conducted to validate and strengthen the reliability of the obtained results. The subjective results reveal that GS methods deliver competitive visual quality compared to NeRF-based approaches, particularly for FF scenes, while maintaining significant advantage in training and synthesis speed. However, GS methods still face challenges in matching the quality of NeRF-based methods, especially for 360º scenes. Among the evaluated GS methods, Scaffold-GS showed the highest overall synthesis quality, SOG excelled in storage efficiency, and LightGaussian achieved the best balance between speed and storage (but with slightly lower quality). Regarding the objective metrics, similarity-based solutions such as FSIM and MS-SSIM were most effective for 360º scenes, whereas ST-LPIPS excelled in FF scenes, leveraging on its tolerance to minor geometric errors and pixel misalignments. However, the study also exposed limitations in current metrics for detecting GS-specific artifacts, highlighting the need for improved metrics tailored to GS-generated content. As such, this paper provides a baseline for further research on GS quality assessment.

# REFERENCES

[1] B. Fei, J. Xu, R. Zhang, Q. Zhou, W. Yang, and Y. He, “3D Gaussian Splatting as New Era: A Survey,” *IEEE Trans. Vis. Comput. Graph.*, pp. 1–20, May 2024.

[2] B. Kerbl, G. Kopanas, T. Leimkuehler, and G. Drettakis, “3D Gaussian Splatting for Real-Time Radiance Field Rendering,” *ACM Trans Graph*, vol. 42, no. 4, Art. no. 4, Jul. 2023.

[3] B. Mildenhall, P. P. Srinivasan, M. Tancik, J. T. Barron, R. Ramamoorthi, and R. Ng, “NeRF: representing scenes as neural radiance fields for view synthesis,” *Commun. ACM*, vol. 65, no. 1, Art. no. 1, Dec. 2021.

[4] Z. Wang, A. C. Bovik, H. R. Sheikh, and E. P. Simoncelli, “Image quality assessment: from error visibility to structural similarity,” *IEEE Trans. Image Process.*, vol. 13, no. 4, pp. 600–612, Apr. 2004.

[5] R. Zhang, P. Isola, A. A. Efros, E. Shechtman, and O. Wang, “The Unreasonable Effectiveness of Deep Features as a Perceptual Metric,” in *Proc. IEEE/CVF Conf. Comput. Vis. Pattern Recognit. (CVPR)*, Salt Lake City, UT, USA, Jun. 2018, pp. 586–595.

[6] T. Lu *et al.*, “Scaffold-GS: Structured 3D Gaussians for View-Adaptive Rendering,” in *Proc. IEEE/CVF Conf. Comput. Vis. Pattern Recognit. (CVPR)*, Seattle, WA, USA, Jun. 2024, pp. 20654–20664.

[7] P. Martin, A. Rodrigues, J. Ascenso, and M. P. Queluz, “NeRF-QA: Neural Radiance Fields Quality Assessment Database,” in *Proc. Int. Conf. Qual. Multimedia Experience (QoMEX)*, Ghent, Belgium, Jun. 2023, pp. 107–110.

[8] P. Martin, A. Rodrigues, J. Ascenso, and M. Paula Queluz, “NeRF View Synthesis: Subjective Quality Assessment and Objective Metrics Evaluation,” *IEEE Access*, vol. 13, pp. 26–41, Dec. 2024.

[9] Y. Xing, Q. Yang, K. Yang, Y. Xu, and Z. Li, “Explicit-NeRF-QA: A Quality Assessment Database for Explicit NeRF Model Compression,” Jul. 18, 2024, *arXiv*: arXiv:2407.08165 [eess.IV].

[10] H. Liang *et al.*, “Perceptual Quality Assessment of NeRF and Neural View Synthesis Methods for Front-Facing Views,” Oct. 2023, *arXiv:2303.15206 [cs.EESS]*: arXiv:2303.15206 [cs.EESS].

[11] Q. Yang, K. Yang, Y. Xing, Y. Xu, and Z. Li, “A Benchmark for Gaussian Splatting Compression and Quality Assessment Study,” in *Proceedings of the 6th ACM International Conference on Multimedia in Asia*, in MMAsia ’24. New York, NY, USA: Association for Computing Machinery, Dec. 2024, pp. 1–8.

[12] Y. Zhang, J. Maraval, Z. Zhang, N. Ramin, S. Tian, and L. Zhang, “Evaluating Human Perception of Novel View Synthesis: Subjective Quality Assessment of Gaussian Splatting and NeRF in Dynamic Scenes,” Jan. 13, 2025, *arXiv*: arXiv:2501.08072 [cs.CV].

[13] J. T. Barron, B. Mildenhall, D. Verbin, P. P. Srinivasan, and P. Hedman, “Mip-NeRF 360: Unbounded Anti-Aliased Neural Radiance Fields,” in *Proc. IEEE/CVF Conf. Comput. Vis. Pattern Recognit. (CVPR)*, New Orleans, LA, USA, Jun. 2022, pp. 5470–5479.

[14] Z. Yu, A. Chen, B. Huang, T. Sattler, and A. Geiger, “Mip-Splatting: Alias-free 3D Gaussian Splatting,” in *Proc. IEEE/CVF Conf. Comput. Vis. Pattern Recognit. (CVPR)*, Seattle, WA, USA, Jun. 2024, pp. 19447–19456.

[15] Z. Fan, K. Wang, K. Wen, Z. Zhu, D. Xu, and Z. Wang, “LightGaussian: Unbounded 3D Gaussian Compression with 15x Reduction and 200+ FPS,” Nov. 12, 2024, *arXiv:2311.17245 [cs.CV]*: arXiv:2311.17245 [cs.CV].

[16] S. Girish, K. Gupta, and A. Shrivastava, “EAGLES: Efficient Accelerated 3D Gaussians with Lightweight EncodingS,” in *Eur. Conf. Comput. Vis. (ECCV)*, Berlin, Heidelberg, Nov. 2024, pp. 54–71.

[17] W. Morgenstern, F. Barthel, A. Hilsmann, and P. Eisert, “Compact 3D Scene Representation via Self-Organizing Gaussian Grids,” in *Eur. Conf. Comput. Vis. (ECCV)*, A. Leonardis, E. Ricci, S. Roth, O. Russakovsky, T. Sattler, and G. Varol, Eds., Berlin, Heidelberg, Nov. 2024, pp. 18–34.

[18] K. Ren *et al.*, “Octree-GS: Towards Consistent Real-time Rendering with LOD-Structured 3D Gaussians,” Oct. 17, 2024, *arXiv:2403.17898 [cs.CV]*: arXiv:2403.17898 [cs.CV].

[19] J. L. Schönberger and J.-M. Frahm, “Structure-From-Motion Revisited,” in *Proc. IEEE/CVF Conf. Comput. Vis. Pattern Recognit. (CVPR)*, Las Vegas, NV, USA, Jun. 2016, pp. 4104–4113.

[20] J. Schönberger, E. Zheng, M. Pollefeys, and J.-M. Frahm, “Pixelwise View Selection for Unstructured Multi-View Stereo,” in *Eur. Conf. Comput. Vis. (ECCV)*, Amsterdam, The Netherlands, Oct. 2016.

[21] A. Knapitsch, J. Park, Q.-Y. Zhou, and V. Koltun, “Tanks and temples: benchmarking large-scale scene reconstruction,” *ACM Trans. Graph.*, vol. 36, no. 4, Art. no. 4, Jul. 2017.

[22] K. Zhang, G. Riegler, N. Snavely, and V. Koltun, “NeRF++: Analyzing and Improving Neural Radiance Fields,” Oct. 2020, *arXiv:2010.07492 [cs.CV]*: arXiv:2010.07492 [cs.CV].

[23] ITU-R Recommendation BT.500-15, “Methodologies for the subjective assessment of the quality of television images,” *Int. Telecommun. Union*, Jun. 2023.

[24] ITU-T Recommendation P.910, “P.910: Subjective video quality assessment methods for multimedia applications,” *Int. Telecommun. Union*, Oct. 2023.

[25] S. Athar and Z. Wang, “A Comprehensive Performance Evaluation of Image Quality Assessment Algorithms,” *IEEE Access*, vol. 7, pp. 140030–140070, Sep. 2019.

[26] J. H. Kim and I. Choi, “Choosing the Level of Significance: A Decision-theoretic Approach,” *Abacus*, vol. 57, no. 1, Art. no. 1, Nov. 2019.

[27] L. Zhang, L. Zhang, X. Mou, and D. Zhang, “FSIM: A Feature Similarity Index for Image Quality Assessment,” *IEEE Trans. Image Process.*, vol. 20, no. 8, Art. no. 8, Aug. 2011.

[28] B. Zhang, P. V. Sander, and A. Bermak, “Gradient magnitude similarity deviation on multiple scales for color image quality assessment,” in *Proc. IEEE Int. Conf. Acoust., Speech Signal Process. (ICASSP)*, New Orleans, LA, USA, Mar. 2017, pp. 1253–1257.

[29] Z. Wang and Q. Li, “Information Content Weighting for Perceptual Image Quality Assessment,” *IEEE Trans. Image Process.*, vol. 20, no. 5, Art. no. 5, May 2011.

[30] E. Larson and D. Chandler, “Most apparent distortion: Full-reference image quality assessment and the role of strategy,” *J. Electron. Imaging*, vol. 19, p. 011006, Jan. 2010.

[31] Z. Wang, E. Simoncelli, and A. Bovik, “Multiscale structural similarity for image quality assessment,” in *Asilomar Conference on Signals, Systems and Computers (ACSSC)*, Pacific Grove, CA, USA, Dec. 2003, pp. 1398-1402 Vol.2.

[32] V. Laparra, A. Berardino, J. Ballé, and E. P. Simoncelli, “Perceptually optimized image rendering,” *J. Opt. Soc. Am. A*, vol. 34, no. 9, Art. no. 9, Sep. 2017.

[33] P. Gupta, P. Srivastava, S. Bhardwaj, and V. Bhateja, “A modified PSNR metric based on HVS for quality assessment of color images,” in *IEEE Int. Conf. Communication and Industrial Application (ICCIA)*, Kolkata, India, Dec. 2011, pp. 1–4.

[34] H. R. Sheikh and A. C. Bovik, “Image information and visual quality,” *IEEE Trans. Image Process.*, vol. 15, no. 2, Art. no. 2, Feb. 2006.

[35] L. Zhang, Y. Shen, and H. Li, “VSI: A Visual Saliency-Induced Index for Perceptual Image Quality Assessment,” *IEEE Trans. Image Process.*, vol. 23, no. 10, Art. no. 10, Oct. 2014.

[36] K. Ding, K. Ma, S. Wang, and E. P. Simoncelli, “Image Quality Assessment: Unifying Structure and Texture Similarity,” *IEEE Trans. Pattern Anal. Mach. Intell.*, vol. 44, no. 5, Art. no. 5, May 2022.

[37] A. Ghildyal and F. Liu, “Shift-Tolerant Perceptual Similarity Metric,” in *European Conference on Computer Vision (ECCV)*, Berlin, Heidelberg: Springer-Verlag, Oct. 2022, pp. 91–107.

[38] R. K. Mantiuk *et al.*, “FovVideoVDP: a visible difference predictor for wide field-of-view video,” *ACM Trans. Graph.*, vol. 40, no. 4, Art. no. 4, Jul. 2021.

[39] Z. Li, A. Aaron, I. Katsavounidis, A. Moorthy, and M. Manohara, “Toward A Practical Perceptual Video Quality Metric,” *Medium*, Apr. 2017.

[40] ITU-T Recommendation P.1401, “Methods, metrics and procedures for statistical evaluation, qualification and comparison of objective quality prediction models,” *Int. Telecommun. Union*, Jun. 2020.

[41] Q. Wu, H. Li, F. Meng, and K. N. Ngan, “A Perceptually Weighted Rank Correlation Indicator for Objective Image Quality Assessment,” *IEEE Trans. Image Process.*, vol. 27, no. 5, pp. 2499–2513, May 2018.